\documentstyle[12pt]{article}

\newcommand{\f}{\phi}

\newcommand{\be}{\begin{equation}}
\newcommand{\ee}{\end{equation}}
\newcommand{\mb}{\bf}

\date{\today}
\title{A  Phase Transition in the Water Coupled
 to a Local External Perturbation }
 \author{Dmitri Volchenkov and Ricardo Lima \\
{\small \em  CNRS, Centre de Physique Theorique,
Luminy Case 907,} \\
{\small \em 13288  Marseille Cedex 09,  France  }\\
{ \small \em E-mail volchen@cpt.univ-mrs.fr,
lima@cpt.univ-mrs.fr}}
\begin{document}
\thispagestyle{empty}
\noindent
 \maketitle
\begin{abstract}
A flux of  ideal fluid coupled to perturbation
is investigated
 by nonperturbative methods of the quantum field theory.
Asymptotic behavior of the flux  coupled to
perturbation turns out to be similar to that of superfluids.
\end{abstract}


\section{Introduction}

The concept of spontaneous symmetry breaking
is an important one in modern theoretical physics.
It is known  in  statistical physics, condensed matter, and nuclear physics,
as well as in relativistic field theory. An  example
of  model  subject to such a phase transition is given
by an  ideal fluid coupled to  local external perturbation, \cite{1}.
For example,  one may consider a pressure pulse
localized in space of inviscid incompressible unbounded fluid
which rises a fluid current with a net momentum flux ${\cal P}.$
Then ${\cal P}$ plays the role of an order parameter, and the
symmetry  is spontaneously broken when ${\cal P}\ne 0.$

Statistical properties of  the {\em symmetrical phase}
of water response for the local external perturbation
 were discussed in \cite{1} in details.
As an  example of such a behavior one imagines  an
eddy risen around  a point wise distinct
 pressure perturbation.
It was shown that  the energy of hydrodynamic  perturbation is confined
within the eddy formed around the  pulse region  and
 demonstrates a power like decay
$\sim  r^{-4} $ outside this region, \cite{1}.

In the present paper we consider the water response in the
case of symmetry broken, {\em i.e.,} when
 ${\cal P}\ne 0$. Since the symmetry breaking
does not affect the scaling properties of the theory, the  universal
quantities for the long-range  asymptotic behavior of the correlation
functions are   the same as in the symmetrical phase.
 We demonstrate that  if  ${\cal P}\ne 0$ the leading order of
long-range water response for  perturbation   is  not  determined by
the scaling degrees of freedom, but by the anomalies risen due  to an
 explicit  symmetry breaking.
   The results achieved
would  give a key  for better understanding of various aspects in
 the  studies of the ocean surface. For example, one can
consider the problem
of generating of  circulation  motions (meanders) close to
ocean currents. The grown up meanders is about to separate from the
 main current
forming the  stable   closed rings
 (which averaged lifetime is about 2-3 years)
of  hundreds of  kilometers in diameter drifting slowly  along the
main current (with an averaged speed of 2-10 cm/s), \cite{MK}-\cite{K}.

The leading order of  correlation functions stationary spectra  in the
fully  developed turbulence  theory of Kolmogorov \cite{Monin} as well as
in the  statistical theory of waves based on
the Zakharov's kinetic equations \cite{MK} can be found from pure
phenomenology in principle.  However, that is not the case for the
 water coupled  to perturbation.
 The matter is  of  presence of redundant
degrees of  freedom connected to perturbation as well as
to hydrodynamic equations themselves.
To get rid of these degrees of freedom we propose  a physically
 relevant hypothesis on the
 mechanism  coupling water to perturbation.

 In \cite{1}  we used the renormalization group (RG)
technique to justify such an additional assumption for the
case of $O(3)-$symmetrical perturbation.
 In the framework of RG method the physical degrees of freedom are to be
 replaced by the scaling degrees which are related to the physical degrees
 through the RG transformations of fields and parameters of the theory.
Since the properties of scaling degrees of freedom are governed by
  a group structure
  (the renormalization group) one investigates them much easier
  then those of
  the origin problem. The results obtained from RG-analysis are
   considered as
   somewhat statistical steady state limit of the physical system.
  The renormalized correlation functions are
distinguished from   their physical analogies only by normalization
conditions, so that they  can be also used for the analysis of
 asymptotic behavior
 of the physical system.  The investigation of
the relevant scaling degrees of freedom which has been
brought about in \cite{1}
 demonstrates that the asymptotic   behavior
  $r>>l_p$ ($l_p$ is the perturbation scale)
of   water response can be  determined unambiguously
if one supposes a coupling
between translation and rotation components of the velocity of fluid.

The stationary spectra can have place for those scale intervals which are
transparent for  currents  of  conserved quantities.
The conserved energy current  from the large-scale region of  pumping
 into the small-scale range of viscous dissipation
allows  one to adjust the well known
Kolmogorov's spectrum in the fully developed turbulence  (the Five
Thirds Law).
 However, the {\em inertial} range of Kolmogorov lies apart  from the
 scale spectrum  of our interest.
 The  transparency interval related to the  enstrophy current
(the squared averaged vorticity)  ${\cal E}$  is exactly
the scale of the  problem considered, though  the enstrophy current
 gives us  the  spectrum of  vorticial
 component of the velocity field only,
\be
{\cal  E}^{1/2}k \simeq A(k),
\label{a}
\ee
where $\mb A$ is the vector velocity potential:
\be
 \mb v (\mb x,t)= -\mb{grad} {\ } \f(\mb x,t)+\mb{rot} {\ } \mb A(\mb x,t),
\label{v}
\ee
and $\f$ is the scalar velocity potential.
However,  the spectrum for  $\f(k)$ is still
unknown from these phenomenological considerations.

In language of the critical phenomena theory the spectrum
(\ref{a}) determines the {\em critical dimension} $\Delta [A]$ of the
field $A$,  $\Delta[A]=1.$ The use of critical dimension allows to compute
the spectrum of any correlation function of the field $A$ by simple
 dimensional counting.
For example,  for the pair correlation function in Fourier representation
 $D_A(k)\equiv
\langle\mb A (\mb k)\mb A (-\mb k)\rangle,$ one has the asymptotics
\be
D_A(k)\sim k ^{\Delta[A]}, \quad \Delta [D_A]=2\Delta [A]-d,
\label{pair}
\ee
where $d$ is the dimension of space.

 Further phenomenological considerations
allow to determine the spectra of quantities which can be measured in
 experiments,
for example, for the   energy as a  function of distance apart
 from the perturbation point,
 $E(r),$ one can obtain:
\be
{\cal E}r^{-4}\sim E(r).
\label{e}
\ee
These results were derived in \cite{1} for the case of $O(3)-$symmetrical
perturbation
and  justified within the framework of RG-approach. Formally,
they are still valid for the case of symmetry broken, ${\cal P}\ne 0.$

The assumption on the  mechanism coupling water to perturbation
 proposed in \cite{1}
allows to fix the spectrum for  $\f$ in the form
\be
\f(k)\sim k^{7/12}.
\label{f}
\ee
The confinement of  energy within the region of $O(3)-$symmetrical
  perturbation could be interpreted as  a kind of {\em short-range} order
(in analogy with  infinite  ferromagnets), but the {\em long-range} order
is suppressed. In case of symmetry broken spontaneously there is
a nontrivial expectation value for $\langle \f\rangle=
\alpha(\mb x)$
(in analogy to the arising of spontaneous magnetization in ferromagnets)
engaged in long-range order with large-scale current  motion.

In a regular way $\alpha (\mb x)$  is to be determined from the
equation of state
\be
(\alpha-\alpha_0)=f(j),
\label{es}
\ee
with some function $f(j)$ calculating usually in the framework of
Feynman graph expansion, \cite{Br}.  However, the power-like
asymptotic solution for (\ref{es}) can be  derived readily from
phenomenology: considering ${\cal P}$ as a new
dimensional parameter of the theory
with symmetry broken, one  obtains  the spectrum $\alpha(k)$ in the form
\be
\alpha(k)\sim k.
\label{fsb}
\ee
Continuing the analogy with ferromagnets, one can conclude that the
ordered phase (of broken symmetry)
water response can be described by a  quantity analogous to the
longitudinal susceptibility
\be
\chi_L=\frac{\partial \alpha}{\partial j},
\label{sus}
\ee
determined by
\be
\chi_L\equiv \int d \mb x {\ } \langle\left[\f(\mb x) - \alpha(\mb x)\right]
\left[\f(0) - \alpha(0)\right]\rangle.
\label{susdef}
\ee
By the way, from the critical phenomena theory point of view
 the main problems of the theory in the non-symmetrical phase are
  to  determine
an explicit form for the function $f(j)$ in (\ref{es}) and  to  justify
the phenomenological result (\ref{fsb}).

The plan of the paper follows:
First, in the second section,
for the convenience of readers
 we briefly reproduce the main result of \cite{1}
on model of ideal unbounded fluid coupled to perturbation
concluding the section by a discussion on the explicit symmetry
breaking. The statistical properties of hydrodynamical system
can be described by a partition function of statistical mechanics with
a classical euclidean action. As a result we derive the classical
action functional designed to describe the long-range
asymptotic behavior of the water  coupled to perturbation.
The relevant functional turns out to be analogous to that one
of the {\em abelian Higgs model} well-known in relativistic field theory
and superconductivity.

 In the next section we observe the basic properties of the theory
 in case of an explicit $U(1)-$symmetry breaking and consider
 the physical consequences for the asymptotic behavior of the
 model. These properties  are dramatically different from those
 demonstrated by the model in the symmetrical case discussed in \cite{1}.
In the latter phase  scaling degrees of freedom were completely
determined by the vector velocity potential $\mb A,$ though
in presence of comparably strong fluid flow the vector velocity
components are confined in the flux and their contributions
are irrelevant for the asymptotic behavior of the system.
 Looking for the  stable stationary
solutions, we obtain an infinite countable set   of such solutions
 distinguished one from the other  by energy gaps. In particular,
 a "ground state" solution can be interpreted as a pure
  laminar  flow  ($\mb A\equiv 0$).

In the fourth section we consider the asymptotic for $\alpha(r)$
in case of the laminar flow; it is provided by a Goldstone
asymptotics arising by an explicit $U(1)-$symmetry breaking
which has place in the real physical system.

To account the contributions of eddies into flux for
"excited states" we construct an instanton solution
for the theory of water coupled to perturbation and find
the interval of validity for the phenomenological result (\ref{fsb})
in the fifth section.

In the Section 6, we discuss the results obtained from the
point of view of dynamical systems theory. We conclude in the last section.

\section{The Model of Ideal Unbounded Fluid
Coupled to Perturbation}

In  \cite{1} it was shown  that after  elimination of  all
redundant degrees of  freedom the problem  considered
 possesses   the  entire $U(1)-$gauge symmetry \cite{gft}.
In case of  $ j\ne 0$ this symmetry turns out to be hidden
by an explicit symmetry breaking term appearing in the effective
action functional.  To  reveal the hidden symmetry of hydrodynamics of
ideal fluid  we  first  suppose  $ j=0$.
Then the equation of hydrodynamics takes the form
  \be
\mb {div}{\ } \mb v(\mb x,t)=0, \quad
p(\mb x,t)= \int {\ }d\mb y \frac{\partial_iv_j(\mb y,t)
\partial_jv_i(\mb y,t)}
{|\mb x-\mb y|}.
\label{ce}
\ee
Here $\mb v(\mb x,t)$ is the velocity of fluid and
$p(\mb x,t)$ is the pressure distribution. The trivial
boundary conditions for the fields at infinity are implied.

The equations (\ref{ce}) do not lead to a hamiltonian in
the usual way, since it is not possible to define
the conjugated moments. Nevertheless, these equations  can be
derived  from the classical Lagrangian ${\cal L}(\mb \varphi)$
formulated in  favor of the scalar and vector velocity potentials, \cite{1}:
\be
{\cal L}(\f, A_i, p) =\frac 12\int d{\mb x}\left[
(\partial p)^2+(\partial \f)^2+\frac 12 F^2+p\partial_iv_j\partial_jv_i +J_iA_i
\right],
\label{lagr}
\ee
where we have introduced the eddy component of velocity
field  as a {\em gauge} invariant tensor
$F^k=(\mb {rot}{\ }\mb A)_k=\partial_iA_j-\partial_jA_i,$
$J_i=\partial_jF^k-\partial_kF^j$ ($i\ne k\ne j$) is the
{\em vorticity} conserved for ideal fluid, and
the tensor  $v_iv_j\equiv (\partial_i\f)(\partial_j\f)+
F^iF^j-\partial _i\f F^j-\partial_j\f F^i$ (the last two terms
are vanished when integrated over $\mb x$ with trivial boundary conditions).

With no coupling to pressure field ($p=0$)
 (\ref{lagr}) is invariant under the following
transformations of fields
\be
\left\{
\begin{array}{l}
A_i(\mb x)\to A_i(\mb x)-\partial_i \Lambda(\mb x), \\
\f(\mb x)\to \f(\mb x)e^{iu\Lambda(\mb x)},
\end{array}
 \right.
\label{gi}
\ee
where $u\Lambda(\mb x)$ is  an arbitrary continuous, differentiable phase function.
The relations (\ref{gi}) express the $U(1)-$gauge symmetry of the
functional (\ref{lagr}) ($U(1)$ is the group of multiplication
by complex numbers). In accordance to the Noether's theorem
(\ref{gi}) relates to two  conserved currents,
 vorticity, defined above,
\be
\partial_iJ_i=0,
\label{cc1}
\ee
and a current related to rotations in the complex plain $U(1)$:
\be
I_i=\f^*\partial_i\f-\f\partial_i\f^*, \quad \partial_iI_i=0.
\label{cc2}
\ee
The statistical properties of mechanical system of an infinite number
of degrees  of freedom can be derived from the partition function
of statistical mechanics $Z= Tr (e^{-S})$ with  somewhat classical
euclidean action functional,
\be
S(\f,A_i,p)=\frac 12 \int\left[ (\partial p)^2+(\partial\f)^2+
\frac 12 F^2 +p\partial_iv_j\partial_jv_i \right]
\label{s}
\ee
The last term in (\ref{s}) does not meet the entire symmetry (\ref{gi}),
since the pressure field as it is included in the action functional contains
somewhat redundant degrees of freedom, we therefore can integrate it over
 in the partition function $Z$. The result of functional integration
 does not depend on $p(\mb x,t)$. This procedure is reduced to elimination
  of the quadratic term proportional to $(\partial p)^2$ from
  (\ref{s}) and the replacement of the $U(1)-$symmetry breaking
  term by
  \be
  \frac 12 \int d{\mb x} {\ } \partial_iv_j(\mb x,t)
  \partial_jv_i(\mb x,t)\int _{V_p}d \mb y \frac
  {\partial_iv_j(\mb y,t)\partial_jv_i(\mb y,t)}
  {|\mb x-\mb y|},
  \label{t}
  \ee
which relates the fluctuations of velocity fields risen by the
perturbative pulse in the perturbed region $V_p$ to those
fluctuations apart from $V_p$.

In \cite{1} we investigated  possible contributions of (\ref{t})
into the action functional considering  the insertions of  various
power like composite operators. It was shown that
the only component   which is important  from the  RG point of view  have to be
 $U(1)$-gauge symmetrical, {\em i.e.,}
 \be
 m^2\f^2
 \label{m2f2}
 \ee
in the first order,
where $m^2$ is somewhat mass parameter (the coefficient of the relevant
RG-invariant composite operator). The use of Ward identities which
 express the $U(1)-$gauge invariancy of the theory allows to demonstrate
 that all other combinations of quadratic operators are ultra-violet
 (UV) finite, {\em i.e.,} the relevant correlation functions do not
 have UV-divergencies, and then they do not  participate in scaling degrees
of freedom.

Instantly close to the region of perturbation the pressure pulse
rises the wave motions with eigenmodes $k>k_0\simeq V_p^{-1/3}.$ Due
to strong nonlinearity of  interaction in  hydrodynamical
equations the eigenmodes of oscillations spread very fast from a
band of order $V_p^{-1/3}$ over the whole spectrum, and various multipole
oscillations of any type are arisen with time. Clearly, the long-range
fluid behavior will depend to some extent on the statistical
properties of wave mode coupling. Following \cite{1} we suppose
the simplest model for the coupling mechanism by inclusion of
the $\varphi^4$-type interaction term into (\ref{s}) with  a wave
modes coupling constant $g>0$. In accordance with the general
critical phenomena approach we note that the accounting of highest
oscillation harmonics, {\em i.e.,} $\varphi^6$, $\varphi^8,$
and so on cannot alter the large-distance asymptotic behavior of
 water response if $g\ne 0.$ Again, since  we are interested in
 $U(1)-$gauge symmetrical term  the only amendments into action
functional  have to be of the form, $\sim g\f^4.$

As a resulting hypothesis  we obtain the effective action
functional to be:
\be
S(\f, A_i)=\frac 12 \int d\mb x \left[(\partial \f)^2+
\frac 12 F^2+m^2\f^2+\frac 13 g\f^4\right]
\label{se}
\ee
 which is designed to  describe the asymptotic properties of water response.

The action (\ref{se}) has some redundant degrees of freedom, the
gauge degrees, with unknown dynamics. As a consequence it
is not renormalizable, and it has no solutions in the massless limit $m^2=0,$
 \cite{Br}.
To construct  a renormalizable theory we  introduce an abelian gauge
geometrical structure:

(i) $\f({\bf x})$ and $\f^*({\bf x})$ are vectors
for $U(1)$ transformations,

(ii) The derivative $\partial_i$ is replaced by the
 covariant derivative $\nabla_i$:
   \be
   \nabla_i=\partial_i+iu_0A_i,
   \label{A7}
    \ee
    where
$u_0$ is the coupling constant of interaction between
the scalar and rotational components of the velocity potential $\varphi$
 (analogous to the electron charge $e$ in electrodynamics).

 (iii) It follows that the curvature tensor
is $iu_0F_{ij}:$
 $$ iu_0F_{ij}=[\nabla_i,\nabla_j]=iu_0(\partial_iA_j-\partial_jA_i).$$

(iv) Since the $U(1)$-gauge group is abelian
 ($\mb A(\mb x,t)$ is a translation invariant), one can write
  the parallel transporter $U(C)$ along
any continuous contour $C$  which is an element of $U(1).$ In
 terms of a line integral:
   \be
    U(C)=\exp\left[-iu_0\oint_C A_i(s) {\ } ds_i\right]
\label{pt}
 \ee
 as a consequence of   vorticity conservation for ideal fluid.
  Thus, the rotational component of   velocity potential just
   carries on the
fluctuations of the scalar potential field $\f({\mb x})$.
  By the way, two solutions for different points $\f({\mb x},t )$
  and $\f ({\mb y},t)$ are related through the parallel
 transporter (\ref{pt}),   where $C$ is an integration path
 connecting the points $\mb x$ and $\mb y$, \cite{gft}.

Now, the gauge degrees of freedom  (correspondent to invariancy
of  velocity with respect to the  $\partial_i\Lambda$-shifts of
vector potential  $A_i$) can be taken into account by the usual
procedure analogous to the Faddeev-Popov quantization \cite{Br}.
In particular, it leads to inclusion of a gauge dependent term into
the action functional,
   \be
  S (A_i,\f)=\frac 12 \int d\mb x {\ } \left[ |\nabla_i\f|^2+ \frac 12 F^2_{ij}
+\zeta^{-1} (\partial_i A_i)^2+m^2\f^2+ \frac 13 g \f^4\right],
  \label{A9}
  \ee
where $\zeta$ is an arbitrary valued ($ \zeta\in [0,\infty) $)
 auxiliary gauge parameter of the theory.
  The model
(\ref{A9}) demonstrates the existence of a statistically
steady state independently of the details of velocity evolution.

In  \cite{1} the model (\ref{A9}) has been investigated in the symmetrical
phase, $m^2>0$. The crucial distinction between symmetry implementation in
the cases of positive and negative signs for $m^2$  lies though
 in the structure of the "ground state" ({\em i.e.,} the expectation
 value  of velocity  potential).

 Suppose, first, that
$\mb A=0$ in (\ref{A9}), then one has the standard model of a scalar unharmonic
oscillator.  For $m=0$ the oscillator is subject to a phase transition.
At the classical level in the symmetrical phase ($m^2>0$) the oscillator
 model describes the fluctuations having the trivial expectation value
of the field, $\langle\f \rangle=0$ (see Fig. 1.a).
 If $m^2<0$, the system allows
an infinite number of  possible expectation values
related to each other by the unitary transformation group $U(1).$
In particular, if  one fixes a phase parameter of the group $U(1)$
under certain physical conditions, then  for the field $\f$ there are
 two possible mean values  (see Fig. 1.b)
\be
\alpha(\mb x)\equiv\langle\f \rangle=\pm
\sqrt{ m^2 /g}=  \alpha_0(\mb x).
\label{alpha}
\ee
 The latter situation is usually referred to as spontaneously broken symmetry.
 Rise of  a net fluid current  from the region of initial perturbation
  into   outside (see Fig.2)  one can treat as a result of spontaneously
symmetry breaking  which can be described by (\ref{A9}) with $m^2<0$.

If we held  $\mb A=0,$ the lagrangian is still invariant under the set  of
$U(1)$-transformations with no gauge section. When $j=0$,
the Goldstone theorem  predicts  the appearance of a massless
degree of freedom corresponding to  unphysical "angular motion" for which
there is no restoring force. The physical interpretation of such a
degree of freedom  would  be the following: the quantity (\ref{sus})
diverges as   $j\to 0, $ {\em i.e.,} an infinitely small initial fluid flow
$j$ risen by perturbation generates  the nontrivial expectation value
for vector potential, $\alpha\ne 0.$

 If we assume that   $j \ne 0,$ then the action (\ref{A9})  has  the
$U(1)$-symmetry  breaking  explicitly by the new term
 \be
-\int d\mb x {\  }  j(\mb x) Re[\f(\mb x,t)],
\label{sbt}
\ee
where $Re[\f(\mb x,t)]$ is the real part of the complex valued field $\f$.
  This symmetry breaking term gives in the first order in $j$
a mass proportional to
$j^{1/2}\partial_iI_i$ (the axial current $I_i$ is no more conserved)
to the unphysical angular degree of freedom correspondent to
rotations in the complex plain ({\em i.e.,} (\ref{sus}) has no
more divergent).

The situation is though  to be changed dramatically if one includes the vorticial
velocity component into consideration ($\mb A\ne 0$).  Due to
so called {\em Higgs mechanism} the angular degree of freedom does not
 produce divergences in (\ref{sus}) even in the zero order in $j$, and
the gauge field $A_i$ acquires a mass without  spoiling the gauge invariance
and renormalizability of the theory, \cite{gft}-\cite{7}.
These ideas which are quite familiar in the weak interactions theory and
superconductivity allow an heuristic interpretation also in hydrodynamics:
In the symmetrical phase ($j=0$)  $\mb A$ plays
the purely transporting role for scalar velocity potential fluctuations from one point to
another in accordance with  (\ref{pt}). When the  symmetry is broken spontaneously,
  $\mb A$ acquires the longitudinal polarization degree of freedom
giving it a mass; as a direct consequence a vector potential field can penetrate
 only exponentially into the fluid flow with a range proportional to
the inverse of the acquired mass. Like a superconductor expels a magnetic field
from its interior, except for a thin layer at the surface over which the field
decreases exponentially, the fluid flow ousts the eddies from its interior
onto the periphery. The microscopic origin of the Higgs phenomenon in
hydrodynamics lies in screening currents of fluid compensating the
external velocity rotational component (see Fig.3).

In \cite{1} it was shown that the scaling  degrees of freedom proportional to $g$
are vanished from the asymptotic behavior of fluid involved into eddy motion. By the way,
considering  $ j=0$ (say, on the periphery skin of a large-scale current),
one can omit the term $g\f^4$ from (\ref{A9}) to obtain
the relevant effective action functional. However, if $j\ne 0$ (within the current),
 from the heuristic point of view it is obvious that the statistical steady state should be
free of coupling to the vector potential $\mb A$, and the solution for $\alpha(\mb x)$
({\em i.e.,} (\ref{fsb})) is to be determined by a Goldstone asymptotics, \cite{Nal}.

\section{The Gauge-Invariant $U(1)$ Theory   of the
  Water Coupled to Perturbation}

 In the present section we develop the heuristic ideas of the preceding one.
We derive the action functional for the theory in case of symmetry
spontaneously broken and investigate its properties.

 Implementing the local transformation
\be
\left\{
\begin{array}{c}
\f(\mb x) \to\left[\alpha(\mb x)+ \f(\mb x)\right]
e^{ iu\Lambda(\mb x)/\alpha(\mb x) }, \\
A_i(\mb x)\to A_i(\mb x) - \frac u{\alpha(\mb x)}\partial_i\Lambda(\mb x)
\end{array}
\right.
\label{gi2}
\ee
to the model action functional relevant to the
system of equations (\ref{ce}), we
fix  the   gauge in such a way that
\be
u\Lambda(\mb x)=2 \pi n \alpha (\mb x), \quad n\in Z.
\label{ug}
\ee
The parameter $u$ which characterize  the  coupling strength of
vorticial and translational velocity components in this gauge
 is related to  a {\em circulation,} $\Gamma$,
\be
\Gamma_n \equiv \oint \mb A{\ } d\mb x = \frac {2n \pi}{u}.
\label{circ}
\ee
Physical degrees of freedom are become clear now (since $\Lambda(\mb x)$ is gauged away
from the theory):
\be
\begin{array}{c}
  S (A_i,\f)=\frac 12 \int d{\!}x \left[ |\nabla_i\f|^2+ \frac 12 F^2_{ij}
+\zeta^{-1} (\partial_i A_i+\sqrt{2}u\alpha\cdot Im[\f])^2+m_{\f}^2\f^2+\right.
\\ \left.+m_A^2A^2 +\frac 13 g \f^4
+u^2_1\f A^2 +\frac 43 g_1 \f^3 - j\f + S(\alpha)
\right],
\end{array}
\label{s2}
\ee
where we have denoted
$u^2_1=u^2\alpha$, $g_1=g\alpha,$ $m^2_A\equiv 2\alpha u$, $m^2_{\f}=2g\alpha^2-m^2,$
and $Im[\f]$ is  an imaginary part of $\f$.
  Comparing (\ref{s2}) and (\ref{A9}), one can see that the vector field $\mb A$
obtains the longitudinal polarization degree of freedom for
which is expressed in (\ref{s2}) as the new mass term $m^2_A A^2$. The longitudinal
components of the vector fields $A_i$ and $Im[\f]$ are ghosts, which both cancel against the
Faddeev-Popov ghost \cite{deWitt} all having the same mass $m_A.$

The behavior of   (\ref{s2}) is in a way very different from (\ref{A9}):
the unphysical imaginary part of the scalar velocity potential $\f$
disappears and the vector field $\mb A$ obtains a mass so that the vorticial
velocity component is short-range only, {\em i.e.,} it is repelled completely
from the flux.

The standard way to illustrate the last sentence is to
demonstrate that the response of the flux for  an elementary vortex
immersed in is equal to zero \cite{Pol}.
Consider the constant  shift transformation (a purely vorticial
constant velocity component)
\be
F_{ij}\to F_{ij}+f_{ij}
\label{st}
\ee
for the gauge section $F_{ij}$ in the disordered theory (\ref{A9}).
Then the partition function $Z[f_{ij}]$ is invariant
under gauge transformations:
\be
f_{ij}\to f_{ij}+\partial_i\lambda_j-\partial_j\lambda_i,
\label{gt2}
\ee
since it can be compensated by the appropriate change $A_i\to A_i+\lambda_i.$
Therefore, in a gauge-invariant theory (\ref{A9}) one has
\be
Z[F_{ij}+f_{ij}]=Z[F_{ij}];
\label{gi3}
\ee
furthermore, one notes that the constant $f_{ij}$
can be removed from (\ref{A9}) by the transformation:
\be
A_i\to A_i +x_jf_{ij}.
\label{tr2}
\ee
In the phase with long-range correlations (\ref{A9}),
the change (\ref{tr2}) is equivalent  to somewhat change of trivial
boundary conditions for the equations (\ref{ce}) at infinity.
In particular, this yields the new term into the partition function
\be
Z[f_{ij}]-Z[0]=\frac{\varepsilon_A}{u^2}\int d\mb x {\ }f^2_{ij},
\label{yr}
\ee
where $\varepsilon_A$ (the Lagrange multiplier) would be some function which has a natural
interpretation as an amplitude of the response of the flux for  an elementary vortex.
Obviously, if the circulation $\Gamma< \Gamma_{0},$ where $\Gamma_{0}$ is some critical
value correspondent to the phase transition point, the vector potential
becomes short-range correlated, and the partition function $Z$
should not depend on $f_{ij}.$
Thus $\varepsilon_A=0$ in the theory (\ref{s2}).

 In case of the vector velocity potential $\mb A$ is strong enough
then, because  of circulation  conservation, it can be allowed in the
fluid flow in the form of narrow flux tubes.
The relation (\ref{circ}) in this context means that there are always an integer number $n$
of such vorticial tubes in the flow, {\em i.e.,} that each of them have
 a source and a sink  (see Fig.4).

 Varying (\ref{s2}) with respect to $A_i$ and $\f^*\f$ with
the boundary condition (\ref{circ}) ({\em i.e.,} fixing  the circulation
$u^{-1}$
to be constant  in ideal fluid), one can easily estimate the energy of
a flux with length $l_f$ as
\be
E\sim u^2m^2_A l_f,
\label{fe}
\ee
which demonstrates the property inherent to a confinement phenomenon.
Note, that (\ref{fe})  could be derived rigorously by considering of
 a Wilson's loop
operator for a point-wise vorticial current,
\be
{\cal J}_i(\mb x)=-iu\oint \delta(\mb x-\mb y) {\ }d\mb y,
\label{vc}
\ee
for (\ref{s2}) (see, for example \cite{2}).

This situation is analogous to that of  superconductors \cite{hooft}: if
electrically charged bosons (Cooper's bound state of an electron pair)
Bose-condensed then there the electric fields become short-range,
 and the magnetic
fields are ousted from the interior. If finally a magnetic field is
admitted inside
a superconductor it can only come in some multiple vortices, never spread
out because of Meissner effect. Following the analogy with
superconductors, one
 can say that the source and the sink vortices confined in the
 fluid flow are kept together in a potential well, and the potential
 is being linearly proportional to their separation.

 As it well known, \cite{7}-\cite{col}, the hidden symmetry begets the
hidden renormalizability: the divergence structure of renormalizable theory
(\ref{A9}) is unaffected by spontaneous symmetry breaking, and the counterterms
needed in  (\ref{s2}) remain those of the symmetrical theory (\ref{A9}).
Consequently, the critical dimensions calculated in \cite{1} for
the quantities in (\ref{A9}) are still valid formally  also for (\ref{s2}).

\section{ Goldstone Asymptotics
of the Water Flux Coupled to Perturbation}

In the present section we construct (\ref{es}) explicitly and
justify the phenomenological result (\ref{fsb}). We shall consider
the theory in the ordered phase assuming that the velocity
field has no vorticial components ({\em i.e.,}
the flux contains no vorticial pairs, $n=0$ in (\ref{ug})). Therefore,
to describe the statistical properties of the system we
can integrate the partition function $Z[\f,\mb A]$ over  $\mb A$
eliminating the vector field $\mb A$ from the theory. The resulting
partition function will depend solely on the scalar velocity
potential, $Z[\f],$ and the relevant action functional
will be identical to those of scalar $\f^4-$theory in the ordered phase
(nonlinear $\sigma-$model):
\be
S=-\frac 12\int d\mb x \left[(\partial\f)^2+\tau \f^2+
\frac{g}{3!}(\f^2)^2-j{\ } Re[\f]\right].
\label{sf4}
\ee
The distinguishing feature of (\ref{sf4}) is the presence of
Goldstone singularities which arise due to an explicit
$U(1)-$symmetry breaking.   The physical origin of these
singularities is following, \cite{Pol}: the scalar velocity potentials
 with different orientation in the complex plane, however,
 correspond to the same  fluid velocity   and though to the
 same energy. The relevant conserved current meets the Ward identity
 in the momentum representation:
\be
k_i\langle I_i(k)\f(-k)\rangle=\langle\f(0)\rangle.
\label{wi}
\ee
Taking $k\to 0$, one concludes that $\langle I_i(k)\f(-k)\rangle$
must be singular in this limit:
\be
\langle I_i(k)\f(-k)\rangle_{k\to 0}=\langle\f(0)\rangle\frac{k_i}{k^2}+
\ldots
\label{sing}
\ee
The general solution for Goldstone asymptotics in (\ref{sf4})
was given in \cite{Nal} for the unbounded theory and then
generalized in \cite{Nal2} to the theory in a half-space.
In particular, the hypothesis \cite{pp} was proven in
\cite{Nal} for any order of $\epsilon-$expansion
($2\epsilon=4-d$) with $j, k\to 0:$
the equation of state (\ref{es}) has the form
\be
(\alpha-\alpha_0)=aj^{1-\epsilon}+bj+\ldots,
\label{ess}
\ee
and the longitudinal susceptibility (\ref{sus}) is to be
\be
\chi_L=a_1j^{-\epsilon} + b_1+\ldots,
\label{sus1}
\ee
the numerical  coefficients $a, b, a_1, b_1$ are specified in
\cite{Nal}. For the transversal susceptibility, $\chi_T\sim j^{-1}$ as
it follows from the Goldstone theorem. Following the discussion in
\cite{Vas}, formulae (\ref{ess}) and (\ref{sus1}) can be interpreted as a
Goldstone scaling (by analogy to critical scaling) for which
$j$ and $k\sim 1/r$ play the role of significant parameters.
The certain Goldstone dimensions $\Delta^G$ belong to $k$, $j$,
and $\alpha(\mb x)|_{j=0}$:
\be
\Delta^G[k]=1,\quad \Delta^G [j]=2,
\quad \Delta^G[\alpha]=d-2.
\label{gr}
\ee
In contrast with critical dimensions (\ref{gr}) are known
precisely as well as the normalized scaling functions of the
simplest correlation functions \cite{Vas}.
The last relation in (\ref{gr}) justifies the result (\ref{fsb}):
performing the inverse Fourier transformation, one obtains
at three dimensions
\be
\alpha(r)\sim \frac 1{r^2}.
\label{fsb1}
\ee

\section{The Instanton Solutions for the Theory of Water
Coupled to Perturbation}

In the previous sections we have considered
the stationary (with no time dependence) stable  solutions
 of (\ref{s2}) which correspond to  the
saddle points (solutions of the  hydrodynamical equations).
 However, in the case
discussed, $j\ne 0,$ the actual hydrodynamical equations,
posses the non-constant solutions also.

In the previous sections we have shown that there is
a countable set of possible stable stationary solutions
(enumerated by an integer number $n$)
for the system of flux coupled to perturbation
distinguished one from the other by the energy gaps (\ref{fe}).
Obviously, the non-constant statistically steady solutions
are related to a specific mechanism of gap generation
(the generation of new pairs of eddies in the flux),
{\em i.e.,} they describe possible transitions
between constant solutions with different $\Gamma_n$ (\ref{circ}).

Another interpretation can be used: since
the source and sink eddies are confined together in the
potential well in the fluid flux, one can consider a tunneling process
of the eddy pair into another potential well.
This tunneling process can be provided by an
{\em instanton } solution \cite{Pol}.
The contribution of instantons into the correspondent partition functional
$Z$ is indeed irrelevant if we are interested in
  relatively short periods of time $t<t_0,$
where $t_0$ is a "tunnelling time". However, for $t>t_0,$
it becomes very large.

Consider the action (\ref{sf4}) in case of the symmetry
broken spontaneously. Classical minima of this action defined from
the equation:
\be
\Delta \f_a-m^2\f_a+\frac g2 \left(\sum_1^2\f^2_b\right)\f_a=j,
\label{e2}
\ee
where $\f_{1,2}$ are the real and imaginary parts of the field $\f$,
$\f=(\f_1+\f_2)/\sqrt{2}$.
 We use the anzatz
\be
\f=\mu (r)e^{i\frac{\Lambda}{\alpha}},
\label{anzatz}
\ee
which gives the equation for $\mu(r)$ in the form,
\be
\mu''-m^2\mu-\frac 2{r^2}\mu +g\mu^3=j.
\label{eq1}
\ee
There exists a solution to (\ref{eq1}) with the properties:
\be
\mu(r\to 0) \to 0, \quad \mu(r\to \infty)\to \alpha_0.
\label{bc}
\ee
The problem of existence and stability of the solution of (\ref{eq1})
with (\ref{bc}) were discussed in \cite{Pol}. The effective asymptotical
solution   is given in the preceding section by the Goldstone asymptotics.
Suppose now that one has introduced a set of vortices, placed at the points
$\mb x_a$ with circulations $\Gamma_a$ into the flux (\ref{sf4}).
The partition function of statistical mechanics $Z$ is then to be
presented in the form (in case of $ \alpha_0\gg u^2$)
\be
Z=Z_0Z_{inst},
\label{pf}
\ee
where $Z_0$ is the standard partition function of the theory (\ref{sf4})
and
\be
Z_{inst}= Tr\left( \exp \left[\frac{\alpha_0}{2u^2}\int d{\mb x}{\ }
\sum_{a\ne b} \frac{2\pi \Gamma_a\Gamma_b L}{|\mb x_a-\mb x_b |}
+ C\sum_a \Gamma^2_a
\right] \right)
\label{zinst}
\ee
($L$ being the size of the flux pattern considered; the second term
is the vortex self-energy). One can see that in case of  large fluid flux
$\alpha_0$ and, consequently, strong confining property
the vortices revolving alternatively  are combined into pairs. Such
pairs have very small influence on the correlation functions and are
irrelevant in case of large $\alpha_0.$ The asymptotics provided by the
instanton solutions is just the same as (\ref{fsb1}).

\section{Discussion from the Point of View of
 Dynamical Systems Theory}

The Navier-Stockes
 equation for an ideal fluid can be replaced  by the relation for
the  pressure field, and the Galilean invariance
 of hydrodynamical equations  is  manifested
as a $U(1)-$gauge invariance (\ref{gi}).

We shall concern with the  phase space (of infinite dimensionality)
relevant to the dynamical system of water coupled to perturbation
and limit ourselves to  a qualitative consideration.
 Picturing the instant states of the system in the phase space,
 we obtain  its phase diagram.
The  stable stationary solutions discussed in \cite{1}
and in the present paper can be interpreted as the attraction regions
  or fixed points of
 trajectories of the system in the phase space.

Consider the manifold of initial conditions  correspondent to the only
solution of  disordered phase (${\cal P}=0$). In \cite{1}
we have shown that it can be realized exceptionally
as an eddy risen around a point-wise distinct
 perturbation. One can imagine this manifold
 as a torus  covered
  by the trajectories tending to some  stable cycle (see Fig.5).

If  we chose  a point  apart from the torus as
the initial condition (for example, the points $A$ or $B$ on the
diagram Fig. 5),  the  system will  leave the vicinity
of the torus and tends to some  region of attraction
which  is closed in a sense  that there are no trajectories
 going out of it. This behavior represents a phase transition in the
language of statistical  theory.
Within  the attraction region the trajectory passes  consequently through an
  infinite set  of  fixed points distinguishing
by the $\Gamma_n$ values (\ref{gi}). Most of these points are
{\em hetreoclinic}, and so that  they  are unstable in a sense
that the smallest deviation from  the certain set of initial
conditions will make the system  trajectory
to jump to some other point. These processes, in principle,
 are to be described  by  the instanton solutions
(see Sec. 5).

This technique would provide us  with  information on the transition
 probabilities  between the particular   {\em heteroclinic } fixed
points. Such a quantity could be naturally interpreted within the context
  of the  dynamical systems theory. Let us surround   each
 {\em heteroclinic }
  fixed point by a ball of radius $\varepsilon$ and  consider a
fixed point  $n_0$ which corresponds to
  the solution   with $\Gamma_{n_0}$ (see Fig. 5). Taking
 $\varepsilon$ to be small enough,
  we  can make the volume of each ball to be finite.  Denote the
volume of  a ball sector starting from
 which   the trajectory of the system drops
 into the  $\varepsilon$-vicinity
 of  the other point, $n_i$,  as $V_   \varepsilon  (n_0\to n_i)$.
Then, one can introduce the quantity
\be
P(n_0\to n_i) =\frac{V_  \varepsilon  (n_0\to n_i)}{\sum_{k, k\ne 0}
V_   \varepsilon  (n_0\to n_k)}
\label{pr}
\ee
which  is analogous to  a transition probability  defined in the
 statistical theory.
If the point  which we have chosen is a {\em homoclinic}
 attractive fixed point, the probability (\ref{pr}) then tends
to zero.  $0<P<1$ for {\em
heteroclinic} points, and  $P=1$ for a repelling point.
We do not know {\em a priori} whether  there are  some {\em homoclinic}
 attractive fixed points in the region of attraction
(see Fig. 5) or there are only the {\em
heteroclinic} points.
 We expect  though   that in case of
$\alpha\gg  u$ there is a degeneracy of solutions in a sense that
 they  are  predicted by
 the Goldstone asymptotics.

\section{Conclusion}

 In a conclusion one can say that the  flux of ideal fluid coupled
to local external perturbation in the region
$r> l_p$ demonstrates asymptotically some properties
similar to those of   superfluids.
In \cite{1} and in the present paper we have considered
 the statistically steady asymptotic solutions of the model
by various nonperturbative techniques of the quantum field theory.
The results on RG-analysis, Goldstone asymptotics, and
instanton-type solutions are, by the way, exact, and they
 demonstrate that  the long-standing hydrodynamical problem
 of water coupled to perturbation, in principle, can be treated as a
 critical phenomenon.

The  relevant  physical system
contains too many redundant degrees of freedom.
To  fix the statistically stable behavior
in the system one needs to  add
some extra assumptions on the character of perturbation
as well as on the character of wave modes coupling.
The model describing such a behavior is subject to
a phase transition managed basically by the
perturbation symmetry. Roughly speaking, the
 symmetry properties of the initial perturbation define whether
the vorticial or translational fluid velocity
components is the most important one for
the long-range asymptotic fluid response.

\newpage

CAPTIONS FOR FIGURES

\vspace{1cm}

FIGURE 1.

a.)
At the classical level in the symmetrical phase ($m^2>0$) the oscillator
 model describes the fluctuations having the trivial expectation value
of the field, $\langle\f \rangle=0$.

b.) If $m^2<0$, the system allows
an infinite number of  possible expectation values
related to each other by the unitary transformation group $U(1).$
In particular, if  one fixes a phase parameter of the group $U(1)$
under certain physical conditions, then  for the field $\f$ there are
 two possible mean values.

\vspace{1cm}

FIGURE 2.

 Rise of  a net fluid current  from the region of initial perturbation
  into   outside.
\vspace{1cm}

FIGURE 3.

To a Higgs phenomenon in hydrodynamics.
When the  symmetry is broken spontaneously,
  $\mb A$ acquires the longitudinal polarization degree of freedom
giving it a mass $m_A$. The fluid flow ousts the eddies from its interior
onto the periphery.

\vspace{1cm}
FIGURE 4.

The kink-type solutions for fluid flow.
The source and the sink vortices confined in the
 fluid flow are kept together in a potential well, and the potential
 is being linearly proportional to their separation.

\vspace{1cm}
FIGURE 5.

a)  The manifold of initial conditions  correspondent to the only
solution of  disordered phase (${\cal P}=0$).   If  we chose  a point
  apart from the torus as an initial condition
 (for example, the points $A$ or $B$),  the  system will
  leave the vicinity of the torus and tends to some  region of
 attraction which  is closed in a sense  that there are no trajectories
 going out of it. This tendency represents a phase transition
in a language of statistical  theory.

b) Surround
  each  {\em heteroclinic }  fixed point by a ball of
 radius $\varepsilon$ and  consider a fixed point
  $n_0$ which corresponds to   the solution
with $\Gamma_{n_0}$.


\begin{thebibliography}{99}
\bibitem{1}
Preprint {\bf CPT-98/P3712}, D. Volchenkov, R. Lima,
{\em Critical Behavior of the Water Coupled to a Local External Perturbation}
\bibitem{Kras}
V.  P.  Krasitskii, {\em J.  Fluid Mech.}  {\bf 272}, 1-20 (1994)
\bibitem{MK} A.S. Monin, V.P. Krasitskii {\em Phenomena on the Ocean Surface},
Gidrometeoizdat, St.-Peterburg, 1985 (in Russian)
\bibitem{K} Kamenkovich V.M., Koshlyakov M.N., Monin A.S.
{\em Synoptic Eddies in Oceans}, Gidrometeoizdat, St.-Peterburg, 1982. (in Russian)

\bibitem{Monin}
 A.S.  Monin, A.M.  Yaglom, {\em Statistical Fluid Mechanics}
 (MIT Press, Cambridge, Mass., 1975), Vol.  2.

\bibitem{Br} E. Brezin, D.J. Wallace, K.G. Wilson, Phys. Rev. B, {\bf 7}, 1, 232 (1973)

\bibitem{2} J.  Zinn-Justin
 {\em Quantum Field Theory and Critical Phenomena}
 (Clarendon, Oxford, 1990)
\bibitem{gft}
M. Guirdy, {\em Gauge Field Theories} (J. Wiley \& Sons, NY, 1991)
\bibitem{7} E.  Leader, E.  Predazzi
 {\em An Introduction to gauge theory and modern particle physics} (Cambridge, 1996)
\bibitem{col} S. Coleman, in {\em Laws of Hadronic Matter}
(ed. by A. Zichichi; Academic Press, 1975)
\bibitem {Nal} Nalimov M. Yu., Theor. and Math. Phys., {\bf 80}, 2, 212 (1989).

\bibitem{deWitt} B.S. de Witt, Phys. Rev. 162, p. 1195-1239 (1967)

\bibitem{hooft} G.'t Hooft,  Vol. 19 in {\em Advanced Series in Mathematical
Physics },
World Scientific, 1994.
\bibitem{Pol} A. M. Polyakov Vol 3, {\em Gauge Fields and Strings} in
{\em Contemporary Concepts in Physics}, Harwood Acad. Publ., 1987.
\bibitem{Nal2} M. Yu. Nalimov, Theor. and Math. Phys., {\bf 102}, 2, 163 (1995)

\bibitem{pp} A. Z. Patashinsky, V. L. Pokrovsky, JTPH {\bf 64}, 4, 1445 (1973)

\bibitem{Vas}Vasil'ev A.N. {\em Functional Methods in the Quantum
Field Theory and Statphysics} (to be published) (in Russian) (1998)
\end{thebibliography}
\end{document}